# DUEL-THRESHOLD PERCOLATION

A.A.Snarskii[1], M.I.Zhenirovskyy[2],

[1] National Technical University of Ukraine (KPI), Kyiv, Ukraine,

[2] Bogolyubov Institute for Theoretical Physics, Kyiv, Ukraine.

ABSTRACT

For thermoelectric, galvanomagnetic and some other effects there may simultaneously exist two percolation thresholds, close to which the effective kinetic coefficients of macroscopically disordered media are critically dependent on the proximity to percolation thresholds, their behavior being described by universal critical indexes.

The presence of percolation threshold in two-phase macroscopically disordered media is due to a simple reason. In case of a large difference between physical properties of phases, for example, when conductivity of the first phase $\sigma_1$ is much larger than that of the second phase $\sigma_1 \gg \sigma_2$, with increasing concentration $p$ of the former prior to percolation threshold $p_c$, the so-called infinite cluster is formed in the medium. Prior to percolation threshold, current «was obliged» to percolation through poorly conducting sections with conductivity $\sigma_2$, whereas above the percolation threshold, the principal current flows along the well conducting phase $\sigma_1$. On passing the percolation threshold, the effective conductivity $\sigma_e$ is drastically increased.

Close to and at the percolation threshold itself, in the so-called critical region, the effective conductivity demonstrates power dependence on proximity to percolation threshold $\tau = (p - p_c)/p_c$ and is characterized by universal critical indexes [1].

$$\sigma_e = \sigma_1 \tau^t, \quad (p > p_c), \quad \sigma_e = \sigma_1 |\tau|^{-q}, \quad (p > p_c), \quad \sigma_e = \left(\sigma_1^q \sigma_2^t\right)^{\frac{1}{t+q}}, \quad |\tau| \leq \Delta, \quad (1),$$

where $\tau = (p - p_c)/p_c$ - is closeness to percolation thresholds, and $\Delta = (\sigma_2/\sigma_1)^{\frac{1}{t+q}}$ - is smearing region (analog of smearing region of second-order phase transition).

The infinite cluster of bad-conductivity phase $\sigma_2$, formed at reduction of concentration up to $\tilde{p}_c = 1 - p_c$, does not affect in any way behavior $\sigma_e$, it remains close smooth. A similar critical behavior, with its critical indexes, takes place for many other effects, namely thermoelectric, galvanomagnetic, 1/f noise spectral density, elastic, etc.

Here, by example of thermoelectric phenomena, we will show that in a three-dimensional case there may simultaneously exist two percolation thresholds $p_c$ and $\tilde{p}_c$, close to which the effective kinetic coefficients behave in a universal and critical way.

Consider a two-percolation system with local kinetic coefficients $a_{ij}$

$$\mathbf{j}_1 = a_{11}\mathbf{E}_1 + a_{12}\mathbf{E}_2, \quad \mathbf{j}_2 = a_{21}\mathbf{E}_1 + a_{22}\mathbf{E}_2, \quad div\,\mathbf{j}_{1,2} = 0, \quad rot\,\mathbf{E}_{1,2} = 0, \tag{2}$$

where $\mathbf{j}_{1,2}$ are thermodynamic flows, and $\mathbf{E}_{1,2}$ are thermodynamic forces.

In particular, (1) includes thermoelectric effects, in this case $\mathbf{j}_1$ is electric current density, $\mathbf{j}_2 = \mathbf{q}/T$, $\mathbf{q}$ is heat flow density, $T$ is temperature, $\mathbf{E}_1$ is electric field strength, $\mathbf{E}_2 = -gradT$, $a_{11} = \sigma$ is conductivity, $a_{12} = a_{21} = \sigma\alpha$, $\alpha$ is Seebeck coefficient, $a_{22} = \kappa/T$, $\kappa$ is thermal conductivity. For simplicity, the case of small thermoelectric figure of merit has been chosen.

Denote local kinetic coefficients $a_{ij}$ in the first and second phases as $A_{ij}$ and $B_{ij}$ respectively. In the standard case, when "conductivity" and "thermal conductivity" of the first phase is much in excess of the second $A_{11} >> B_{11}$ and $A_{22} >> B_{22}$ effective kinetic coefficients $a_{ij}^e$, averages connecting by definition on volume thermodynamic streams and fields behave in the critically close close $p_c$. However, there exists (at least hypothetically) such a case when $A_{11} >> B_{11}$, but $A_{22} << B_{22}$. In this case a drastic change in the effective kinetic coefficient $a_{11}^e$ occurs, like before, when passing through $p_c$, and $a_{22}^e$ when passing through $\tilde{p}_c$, of that concentration when infinite second-phase cluster originates in the medium (if to change concentration from 1). Crosswise effective coefficients $a_{12}^e$ and $a_{12}^e$ in this case behave critically in the vicinity of both the first and the second percolation threshold.

Effective kinetic coefficients of a two-percolation two-phase system (1) can be found knowing the expressions for effective conductivity in a one-percolation system, for example, in a critical area close to percolation threshold.

Knowing (2), one can use the isomorphism method [2] to find concentration dependences of $a_{ij}^e$ in the vicinity of both $p_c$ and $\tilde{p}_c$ and define critical indexes. According to [2], the effective kinetic coefficients $a_{ij}^e$ are expressed through the effective conductivity of a single-percolation medium as follows $\sigma_e = \sigma_1 f\left(p, \sigma_2/\sigma_1\right)$

$$a_{ij}^e = \frac{\left(\mu A_{ij} - B_{ij}\right)f\left(p,\lambda\right) + \left(B_{ij} - \lambda A_{ij}\right)f\left(p,\mu\right)}{\mu - \lambda}, \tag{3}$$

where

$$\mu = \frac{-c + \sqrt{c^2 - 4\det \mathbf{A} \det \mathbf{B}}}{2\det \mathbf{A}}, \quad \lambda = \frac{-c - \sqrt{c^2 - 4\det \mathbf{A} \det \mathbf{B}}}{2\det \mathbf{A}}, \quad c = A_{22}B_{11} - A_{11}B_{22} + 2A_{12}B_{12}, \quad (4)$$

and it is taken into account that the arrays of local kinetic coefficients $\mathbf{A}$ and $\mathbf{B}$ are symmetrical.

Concentration behavior of the effective kinetic coefficients in the entire range of second-phase concentration - $p$ can be obtained using for $\sigma_e$ of a single-percolation system the approximation of self-coordinated Bruggemann field [3]

$$\frac{\sigma_e(p, \sigma_2/\sigma_1)}{\sigma_1} = f(p, \sigma_2/\sigma_1) = \frac{1}{4}\left[3p - 1 + (2 - 3p)\frac{\sigma_2}{\sigma_1} + \sqrt{\left[3p - 1 + (2 - 3p)\frac{\sigma_2}{\sigma_1}\right]^2 + 8\frac{\sigma_2}{\sigma_1}}\right], \quad (5)$$

The presented concentration dependences $a_{ij}^e$ (Fig.1) clearly demonstrate that "conductivity" $a_{11}^e$ changes drastically at $p_c = 1/3$, "thermal conductivity" $a_{22}^e$ at $\tilde{p}_c = 2/3$, and effective thermoelectric coefficient $a_{12}^e / a_{11}^e$ both at the first percolation threshold $p_c$ and at the second $\tilde{p}_c$.

To determine critical indexes, it is necessary to employ expressions for $\sigma_e$ of a single-percolation system, valid close to percolation threshold (2). In this case for effective thermoelectric coefficient $a_{12}^e / a_{11}^e$ we get

$$\frac{a_{12}^e}{a_{11}^e} \sim \begin{cases} \tau^{-t} &, p > p_c \\ |\tau|^q &, p < p_c \end{cases} \quad \frac{a_{12}^e}{a_{11}^e} \sim \begin{cases} \tilde{\tau}^{-q} &, p > \tilde{p}_c \\ \tilde{\tau}^t &, p < \tilde{p}_c \end{cases}, \quad (6)$$

where, by analogy $\tau$, $\tilde{\tau} = (p - \tilde{p}_c)/\tilde{p}_c$ - is closeness to percolation thresholds $\tilde{p}_c$.

Thus, with chosen values of local kinetic coefficients, the effective coefficients behave critically in the vicinity of both $p_c$ and $\tilde{p}_c$ percolation thresholds.

Galvanomagnetic problem is much more complicated, in the general case it is not reduced to conductivity problem. In two partial cases [2] it can be reduced to electric conductivity problem (without the Hall components) of the anisotropic medium, the latter being rather complicated by itself. Here analysis of a galvanomagnetic problem will be made in the approximation of a self-coordinated field, which is extended to galvanomagnetic effects in [4].

Local conductivity tensor in magnetic field **H** for each of the phases is of the form

$$\hat{\sigma}_i = \begin{pmatrix} \sigma_{si} & \sigma_{ai} & 0 \\ -\sigma_{ai} & \sigma_{si} & 0 \\ 0 & 0 & \sigma_{zi} \end{pmatrix}, \quad \mathbf{H} \parallel 0z, \quad i=1,2, \qquad (7)$$

where later on for components the well-known model dependences will be assumed

$$\sigma_s = \frac{\sigma}{1+\beta^2}, \quad \sigma_a = \frac{\sigma\beta}{1+\beta^2}, \quad \sigma_z = \sigma, \quad (\beta = \mu H/c), \qquad (8)$$

where $\mu$ is carrier mobility, $c$ is velocity of light.

Critical behavior near to the second threshold will occur if, for example, a diagonal component of the first phase $\sigma_{s1}$ is much in excess of the second $\sigma_{s1} >> \sigma_{s2}$, and for component "along" the magnetic field the inverse relationship $\sigma_{z1} << \sigma_{z2}$ holds.

According to [4], the self-coordination equation for the effective conductivity tensor $\hat{\sigma}_e$ is of the form

$$(\hat{\sigma}_e - \hat{\sigma}_1)\hat{\Gamma}(\hat{\sigma}_1,\hat{\sigma}_e)p + (\hat{\sigma}_e - \hat{\sigma}_2)\hat{\Gamma}(\hat{\sigma}_2,\hat{\sigma}_e)(1-p) = 0, \qquad (9)$$

where

$$(\hat{\Gamma}^{-1})_{\alpha\gamma} = \delta_{\alpha\gamma} - \sum_{\beta} n_{\alpha\beta} \frac{(\hat{\sigma}_e - \hat{\sigma}_i)_{\beta\gamma}}{\sqrt{\sigma^e_{\alpha\alpha}\sigma^e_{\beta\beta}}}, \quad i=1,2. \qquad (10)$$

Here for spherical inclusions $n_{\alpha\beta} = \delta_{\alpha\beta}/3$.

Fig.2 shows concentration dependences of the effective Hall coefficient (the left-side axis on logarithmic scale) and the nondiagonal component $\sigma^e_a$ derived from the numerical solution of system (9). Both these values, as is evident from Fig.2, change their behavior both nearby the first- $p_c$ and the second - $\tilde{p}_c$ percolation thresholds. If, however, the condition $\sigma_{z1} << \sigma_{z2}$ is replaced by the inverse one, the second percolation threshold will "disappear".

The appearance of the second percolationthreshold is also possible in thermogalvanomagnetic effects.

Note in conclusion that percolation threshold $p_c$ is not a universal value. Thus, for instance, in the case of net problems, for a cubic net $p_c \approx 0.3$, and for a "diamond"-structured net the percolation threshold is considerably larger - $p_c \approx 0.4$. Therefore, a situation is possible when (at least for net problems) $p_c$ and $\tilde{p}_c = 1 - p_c$ will draw close together so much that critical areas will be overlapped. In this case one can no longer use standard percolation relationships of (2) type, and the appearance of new both concentration and field dependences of the effective kinetic coefficients is possible.

**FIGURES**

Fig.1a. Concentration dependence of effective coefficient "thermoelectromotive" - $a_{12}^e / a_{11}^e$.

Fig.1b. Concentration dependence of effective coefficient "conductivity" $a_{11}^e$ (dashed line) and $a_{22}^e$ "heat conductivity" (solid line).

Chosen the next parameters:
$a_{11} = 5 \cdot 10^6$, $a_{12} = a_{21} = 0.01$, $a_{22} = 8.92 \cdot 10^{-4}$, $b_{11} = 3.207 \cdot 10^4$, $b_{12} = b_{21} = 20 \cdot 10^2$, $b_{22} = 320.701$ (arbitrary unit).

Fig.2. Concentration dependences of effective Hall coefficient - $R_e = \frac{1}{H} \cdot \frac{\sigma_a^e}{\left(\sigma_s^e\right)^2}$ (solid line) and nondiagonal component of conductivity's tensor - $\sigma_a^e$ (dashed line). Chosen the next parameters: $\sigma_1 = 1$, $\sigma_2 = 10^3$, $\beta_1 = 1$, $\beta_2 = 10^2$ (arbitrary unit).

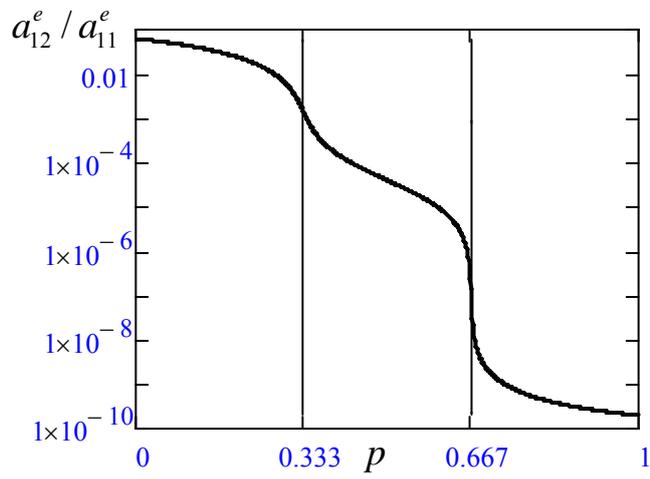

Fig.1a

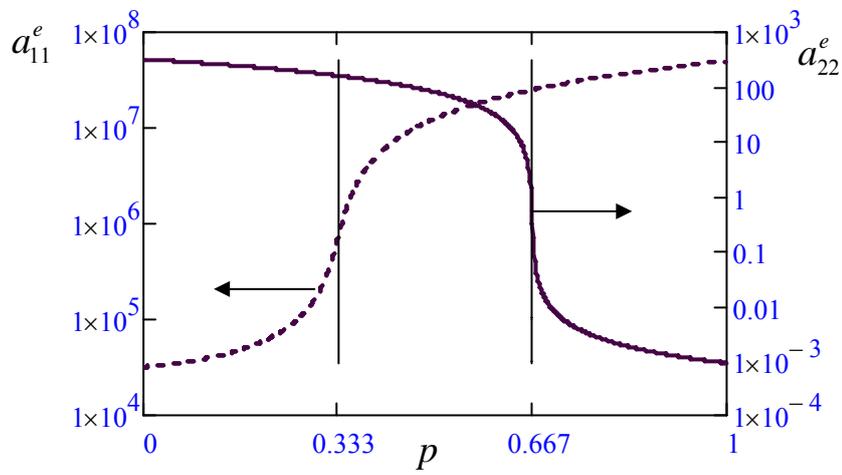

Fig.1b.

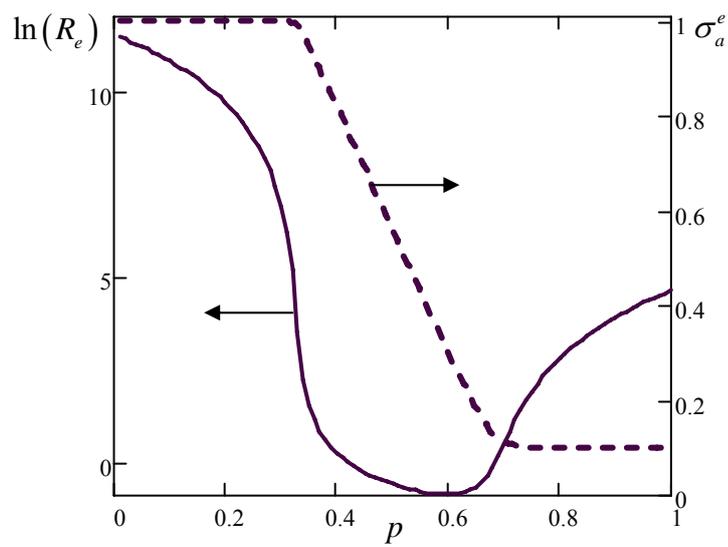

Fig.2.